\documentclass[10pt,conference]{IEEEtran}
\usepackage{amssymb,amsmath}
\usepackage{commath}
\usepackage{subfigure}
\usepackage{cite}
\usepackage{url}
\usepackage{braket}
\usepackage{nicematrix}
\usepackage{longtable}
\usepackage{lipsum}
\usepackage{graphicx}
\usepackage{adjustbox}
\usepackage{rotating}
\usepackage{booktabs,ctable}
\usepackage{textcomp}
\usepackage{multirow}
\usepackage{epstopdf}
\usepackage{verbatim}
\usepackage[T1]{fontenc}
\usepackage{float}
\usepackage{mathabx}
\usepackage{authblk}
\usepackage{epsfig}
\usepackage{algpseudocode}
\usepackage{balance}
\usepackage{multicol}

\usepackage{soul}
\usepackage{nicefrac}
\usepackage[euler]{textgreek} 
\usepackage [latin1]{inputenc}

\makeatletter 
\DeclareRobustCommand*\textsubscript[1]{%
          \@textsubscript{\selectfont#1}}
        \def\@textsubscript#1{%
          {\m@th\ensuremath{_{\mbox{\fontsize\sf@size\z@#1}}}}}
\makeatother

\newcommand{\mytilde}{\raise.17ex\hbox{$\scriptstyle\mathtt{\sim}$}}

\usepackage{graphics}
\usepackage{enumitem}

\begin{document}
\IEEEoverridecommandlockouts 

\title{\huge{Power Delivery for Ultra-Large-Scale Applications on Si-IF}
\vspace{-0.1in}
}
\setcounter{Maxaffil}{1} 
\author[1]{Yousef Safari, Anja Kroon, and Boris Vaisband}
\vspace{-0.4in}
\author{Yousef Safari, Anja Kroon, and Boris Vaisband \vspace{0.15in}\\
	The Heterogeneous Integration Knowledge (THInK) Team\\ 
	Department of Electrical and Computer Engineering,
	McGill University, Montreal, QC H3A 0E9, Canada \\

 	$[$yousef.safari@mail.mcgill.ca$]$\vspace{-0.2in}}

\maketitle
\begin{abstract}
In recent years, with the rise of artificial intelligence and big data, there is an even greater demand for scaling out computing and memory capacity. Silicon interconnect fabric \mbox{(Si-IF)}, a wafer-scale integration platform, promotes a paradigm shift in packaging features and enables ultra-large-scale systems, while significantly improving communication bandwidth and latency. Such systems are expected to dissipate tens of kilowatts of power. Designing an efficient and robust power delivery methodology for these high power applications is a key challenge in the enablement of the Si-IF platform. Based on several figure-of-merit parameters, an efficient power delivery methodology is matched with each of three candidate applications on the Si-IF, namely, artificial intelligence accelerators, high-performance computing, and neuromorphic computing. The proposed power delivery approaches were simulated and exhibit compatibility with the relevant ultra-large-scale application on Si-IF. The simulation results confirm that the dedicated power delivery topologies can support ultra-large-scale applications on the SI-IF.
\end{abstract}
\vspace{-0.05in}
\vspace{-0.05in}
\section{Introduction}
Over the last few decades, system-on-chip (SoC) integration has been the mainstream integration approach in the semiconductor industry for high-performance applications. Recently, Cerebras, a wafer-size compute system has been demonstrated \cite{CR14,CR15}. Cerebras provides a significant increase in compute and memory capacity, but requires fabrication at extremely high yield and does not support heterogeneity. 

Silicon interconnect fabric (Si-IF) is a heterogeneous integration platform and a promising solution to address the challenges of SoCs \cite{CR6}. The Si-IF replaces the complex conventional packaging with a single-layer integration hierarchy. In this technology, dielets (small unpackaged dies) are attached to a Si substrate that serves as the package and the printed circuit board (PCB). The passive Si substrate includes interconnects for signaling and power delivery. The Si-IF is a system-on-wafer that supports the integration of heterogeneous dielets fabricated using disparate technologies, materials, and processes~\cite{CR6,CR7}. 

Power delivery and thermal management are key challenges in wafer-scale systems, especially in assemblies of high power density dielets \cite{CR8}. That said, utilizing Si (relatively high thermal conductivity of 149~W/mK) as the substrate and package, reduces the thermal challenge within the Si-IF platform, as compared to other wafer-scale integration approaches. 

\begin{figure}[t]
	\centering
	\includegraphics[width=1\linewidth,trim=0cm 0cm 0cm 4cm,clip=false]{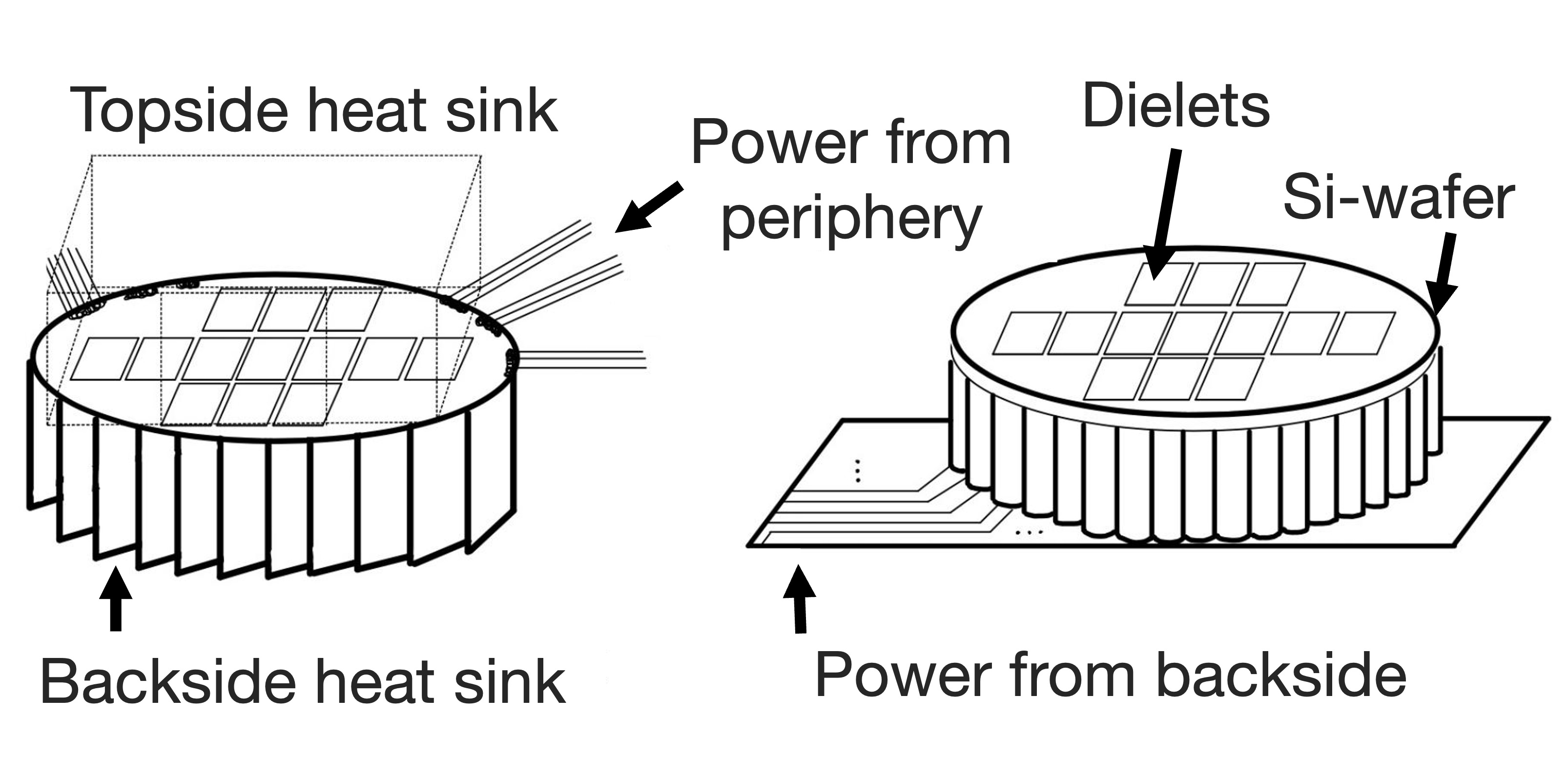}
	\caption{Schematics of the periphery and backside power delivery approaches on the Si-IF.}
	\label{fig_back+periphery}
	\vspace{-0.2in}
\end{figure}

As shown in Figure~\ref{fig_back+periphery}, two general approaches for power delivery on Si-IF are considered, (i)~from connectors at the periphery of the wafer, and (ii)~from the backside of the platform. In the first approach, converters and regulators are placed at the periphery of the wafer, and the current will be delivered to the dielets using the Si-IF horizontal interconnects \cite{CR10}. Although cheaper and easier to implement, peripheral power delivery suffers from significant resistive losses as well as reduction of on-Si-IF area dedicated to functional dielets, and therefore only compatible with low-power applications. Alternatively, in the second approach, converters are placed on a dedicated external PCB or at the backside of the Si-IF, and the current will be delivered to the dielets using through-wafer vias (TWVs) \cite{CR9,CR27}. Here, high-power applications can be supported while requiring high fabrication cost and complexity.

Three principal wafer-scale applications, including artificial intelligence (AI) accelerators, neuromorphic computing, and high-performance computing (HPC), are considered in this paper. For each application, a commercial SoC-based product is chosen and modeled using optimal-size dielets for integration on the Si-IF. Four power delivery topologies (based on either peripheral or backside power delivery) are simulated, modeled, and compared with respect to several figure of merit (FOM) parameters. Specifically, resistive and inductive voltage drop, resistive and inductive power loss, area, and fabrication challenges. 

The rest of the paper is composed of the following sections. The structure of the Si-IF and the proposed power delivery topologies are presented in Section~\ref{section_2}. In Section~\ref{section_3}, the selected commercial products for each application of interest are introduced. Simulation results, comparison of the proposed topologies, and related discussions are provided in Section~\ref{sec_simulation_results}. Finally, some concluding remarks are offered in Section~\ref{sec_conclusions}. 
\section{Si-IF and Power Delivery Topologies}
\label{section_2}
The Si-IF is a wafer-scale chiplet-based platform that supports a small inter-dielet spacing (<~100~\textmu m) and 
a small vertical (between dielet and platform) interconnect pitch (<~10~\textmu m). These parameters allow an ultra-large heterogeneous system to be integrated with high density on a 300~mm wafer~\cite{CR6}.

\begin{figure}[t]
	\centering
	\includegraphics[width=1\linewidth]{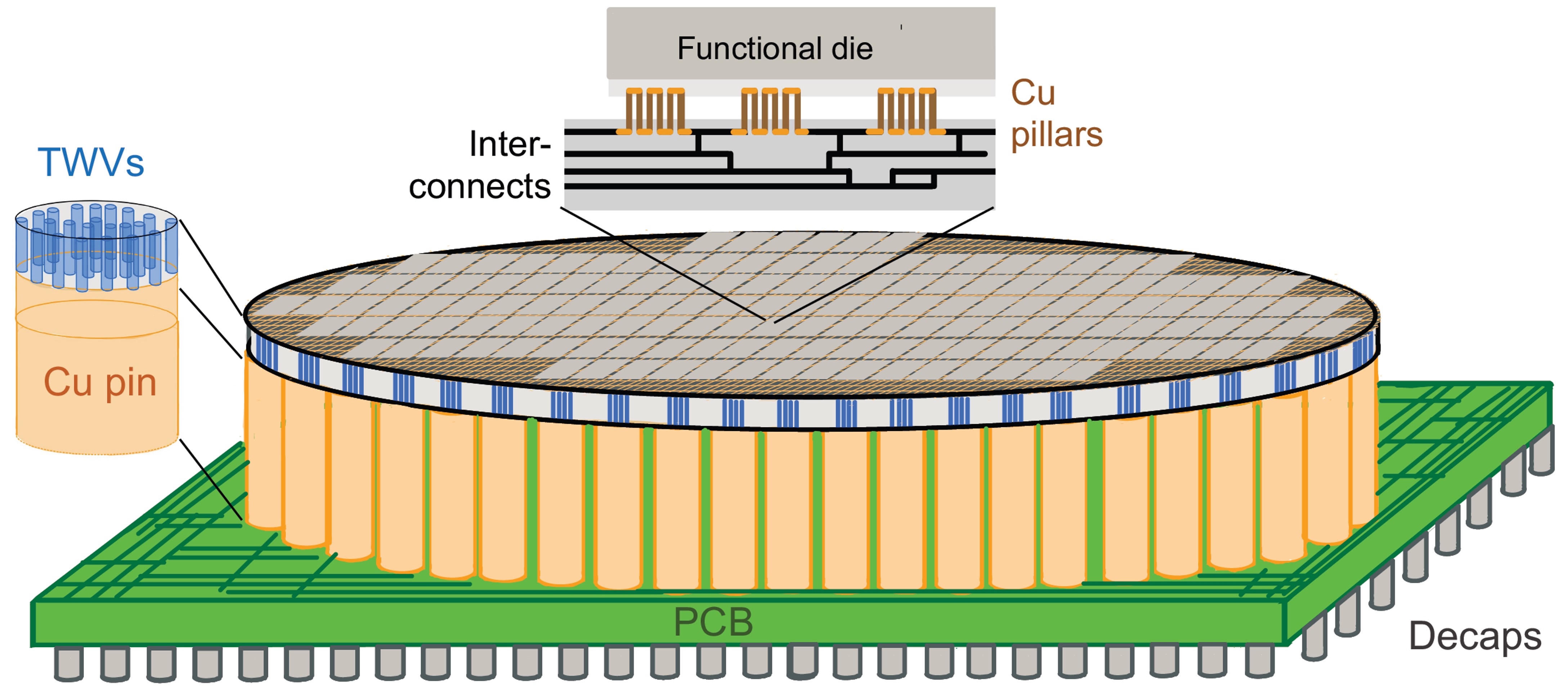}
	\caption{A schematic of the Si-IF platform.}
	\label{fig_si_if}
	\vspace{-0.25in}
\end{figure}
As shown in Figure~\ref{fig_si_if}, the structure of the Si-IF consists of copper (Cu) pins, Si wafer, TWVs, interconnects, Cu pillars, Cu pads, and dielets. Note that the Cu pins, TWVs, and PCB are optional components of the platform for backside power delivery and cooling purposes, \emph{e.g.,} flash cooling~\cite{CR20}. Dielets on the Si-IF are either functional dielets (FDs) or utility dielets (UD). UDs are critical nodes in the network that enable global communication, power conversion and management, synchronization, processing and memory capabilities, redundancy allocation, and test of the Si-IF \cite{CR12}. UDs on the Si-IF platform are similar to routers within a network on chip (NoC) architecture. 
Additional properties of the Si-IF are described in detail in ~\cite{CR6}. Typical dimensions of features on the Si-IF platform are listed in Table~\ref{table_si_if_feat}. 
\begin{table}[b]
	\vspace{-0.1in}
	\caption{Typical feature dimensions of the Si-IF platform~\cite{CR1,CR11}.}
\label{table_si_if_feat}
	\vspace{-0.05in}
	\renewcommand{\arraystretch}{1}
	{\begin{center}

			\begin{tabular}{ll}
				\toprule
			\textbf{Feature}    &   \textbf{Value}	\\	
				\toprule
				
                Optimal dielet area & 1--100~mm\textsuperscript{2} \\
                Inter-dielet pitch & 100~\textmu m \\
                Maximum substrate area & 70,685 mm\textsuperscript{2}\\
                Contact pad area & 20~\textmu m\textsuperscript{2} \\
                Cu pillar diameter/height/pitch & 5/5/10~\textmu m\\
                Cu pin diameter/height/pitch & 1.5/20/3~mm \\
                TWV diameter/height/pitch & 100/500/200~\textmu m \\
                Si substrate diameter/thickness & 300/0.5~mm\\
                Interconnect width/thickness/pitch & 2/2/4~\textmu m \\
                Number of interconnect layers & 2--4 \\

				\bottomrule
			\end{tabular}
		\end{center}
	}
	\vspace{-0.07in}
\end{table}

\begin{figure*}[t]
    \centering
    \subfigure[]{
    \includegraphics[scale=0.2]{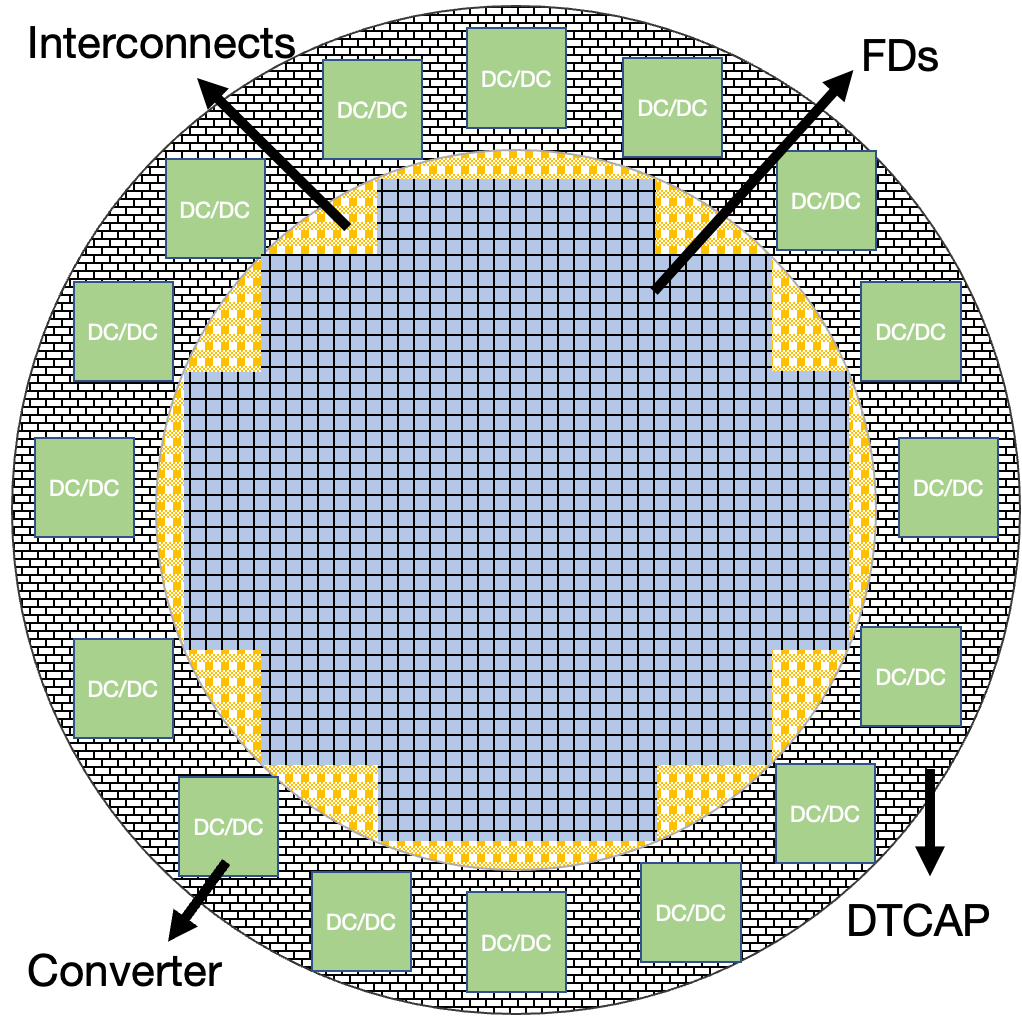}
    \label{top_PT}
    }
    \subfigure[]{
    \includegraphics[scale=0.05]{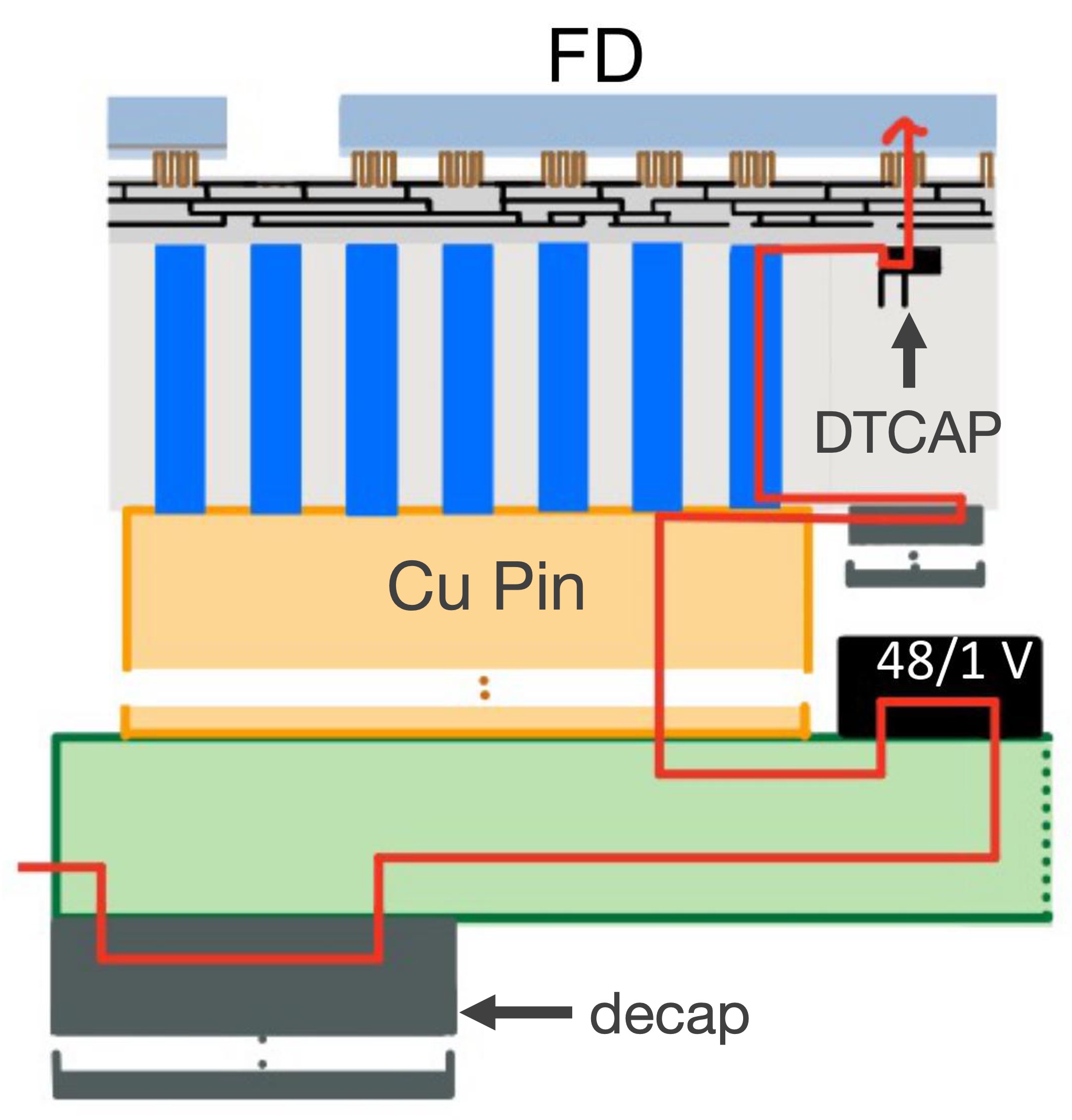}
    \label{top_BT1}
    }
    \subfigure[]{
    \includegraphics[scale=0.05]{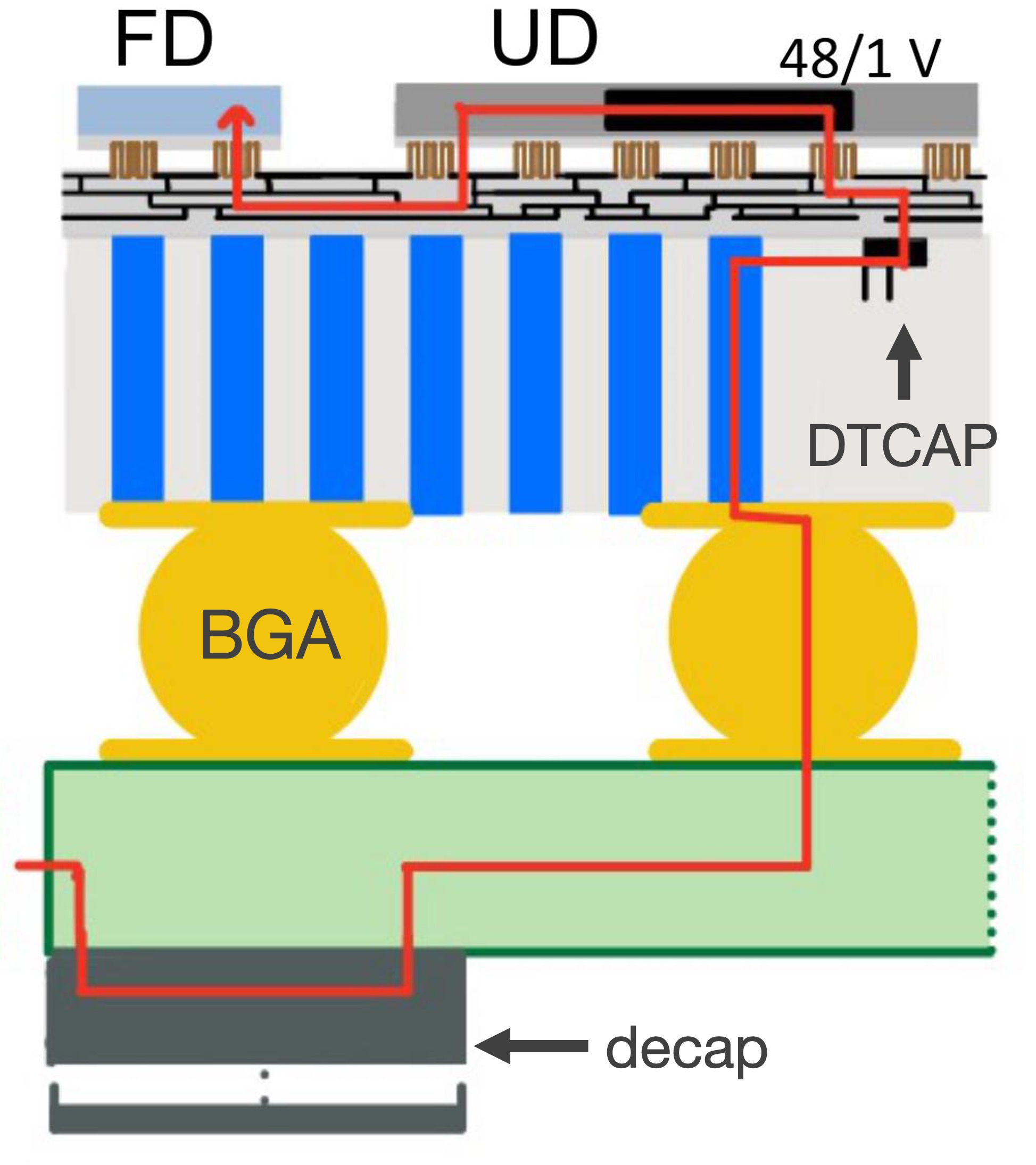}
    \label{top_BT2}
    }
    \subfigure[]{
    \includegraphics[scale=0.055]{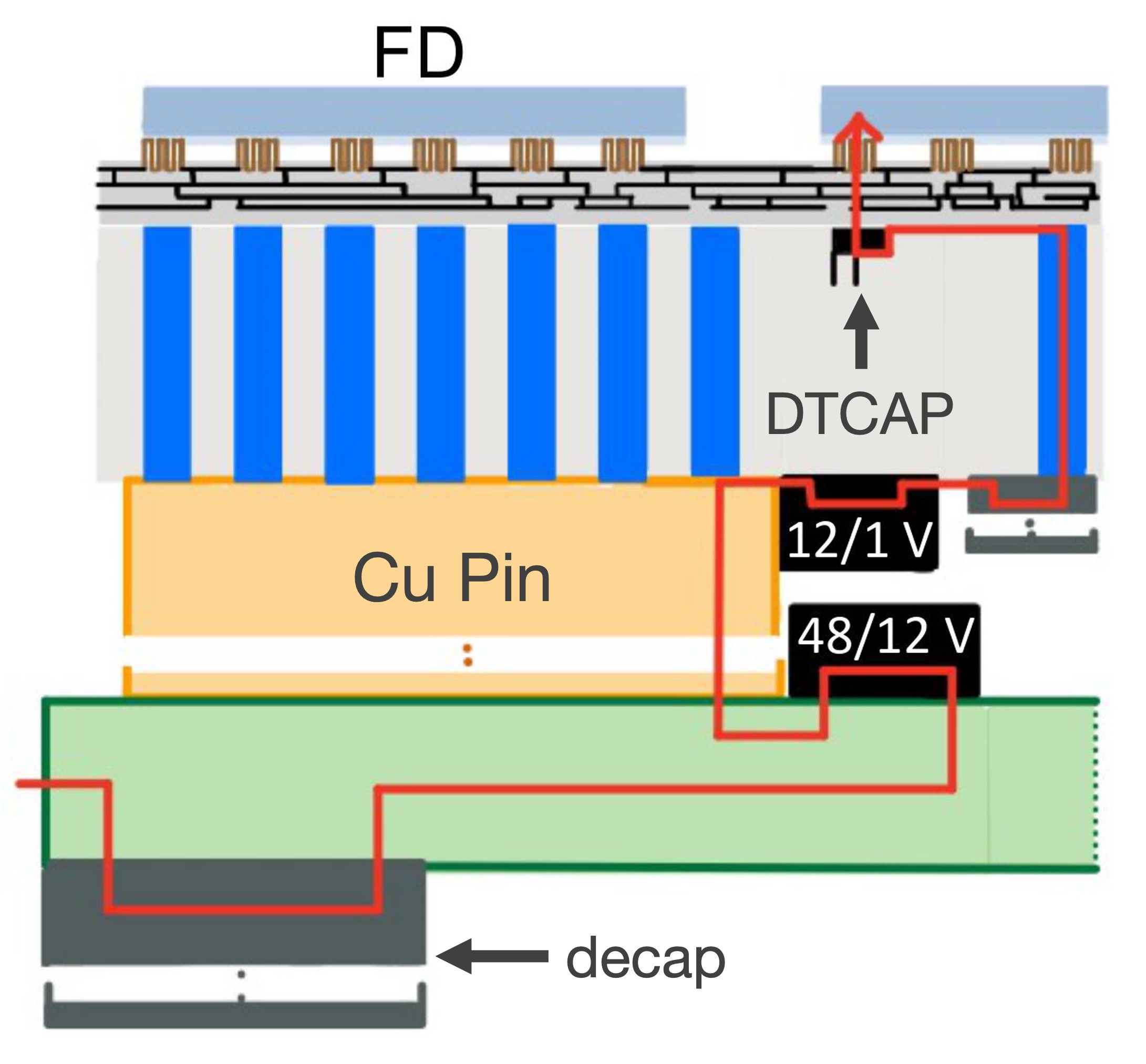}
    \label{top_BT3}
    }    
    
    \caption{Schematics of power delivery topologies: (a) PT, (b) BT1, (c) BT2, and (d) BT3.} 
    \label{topologies}     
    \vspace{0.05in}
\end{figure*}

\begin{figure}[htbp]
\centering
    \subfigure[]{
    \includegraphics[scale=0.05]{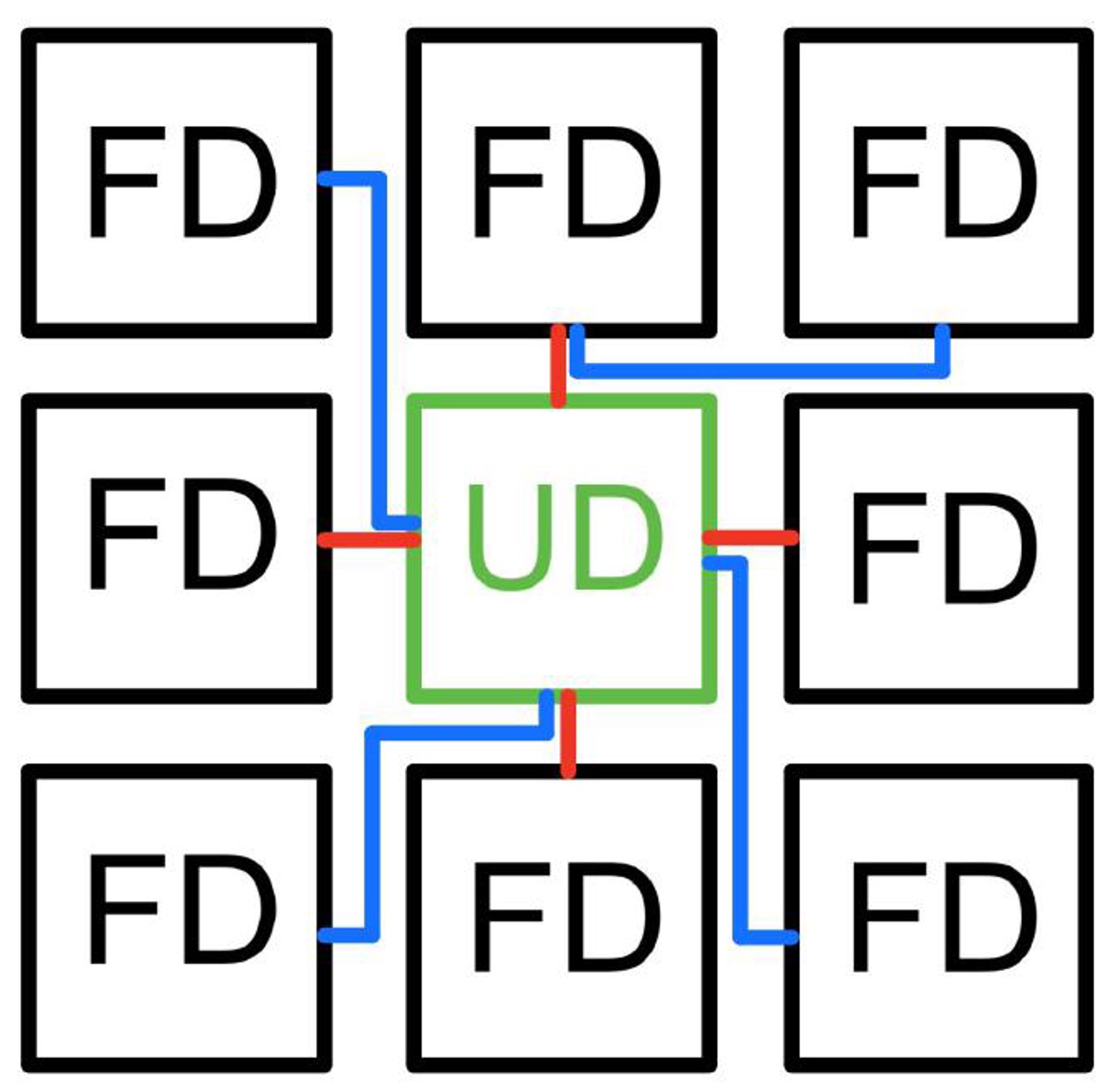}
    \label{fig-coupon-8-1}
    }
    \subfigure[]{
    \includegraphics[scale=0.04]{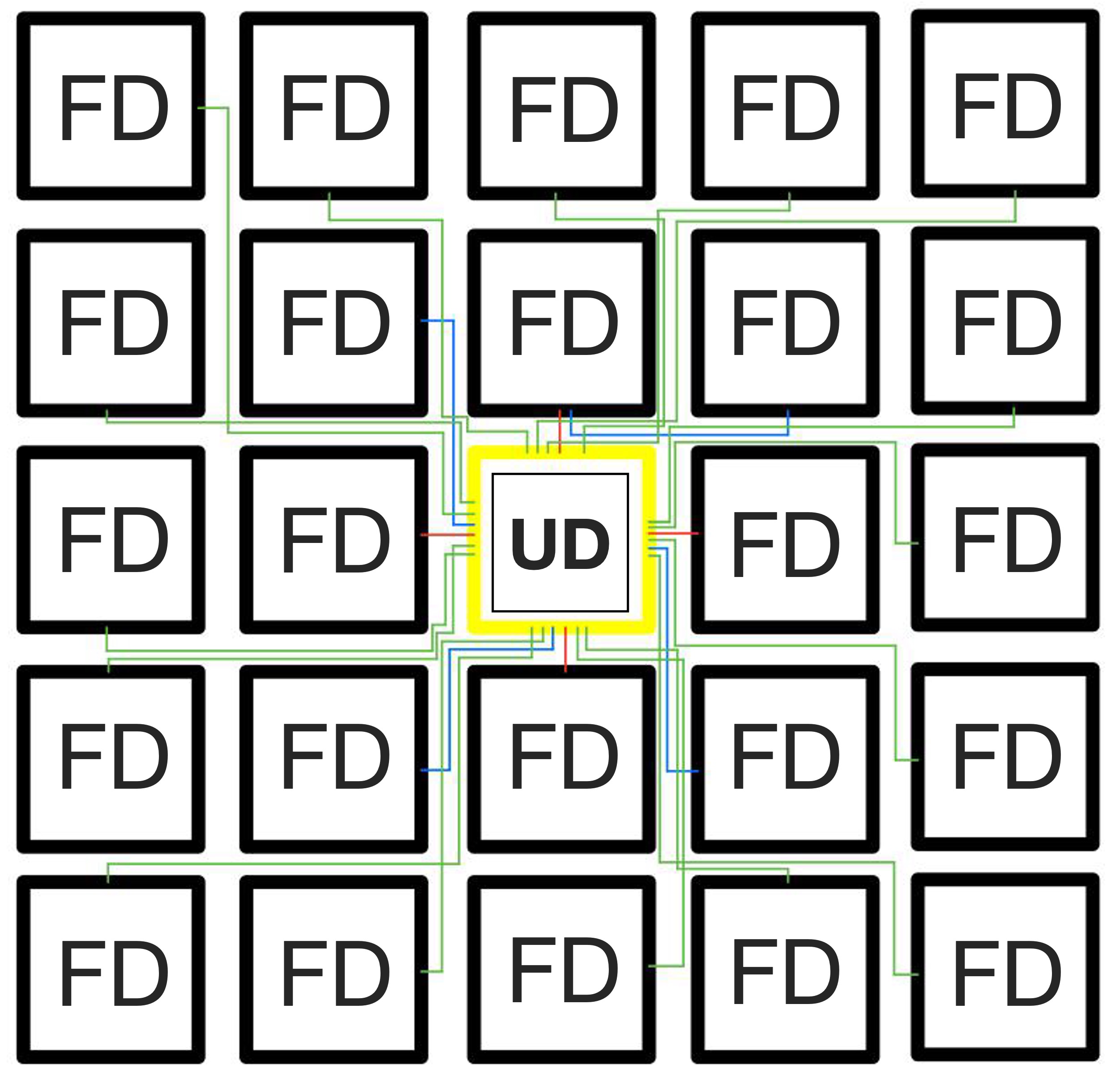}
    \label{fig-coupon-24-1}
 
    }
    
    \caption{Two tile arrangements for BT2. (a)~8--1~tile structure, and (b)~24--1 tile structure.}
	\label{fig_coupon}
	\vspace{-0.2in}
\end{figure}

Schematics of the power delivery topologies under evaluation, including one peripheral (PT) and three backside (BT1, BT2, and BT3) topologies, are shown in Figure~\ref{topologies}. As shown in Figure~\ref{top_PT}, PT includes a ring at the periphery of the wafer which is dedicated to the converters, regulators, and required passives to reduce the high input voltage (48~V) to point-of-load (POL) voltage (48/POL). The current, in this case, will be delivered to the dielets at POL voltage through on-Si-IF interconnects and Cu pillars. It is assumed that, in the PT topology, 30\% of the wafer area is occupied by converters, regulators, and required passives \cite{CR10}. BT1 is a low-voltage high-current topology that includes a 48/POL converter on a dedicated external PCB (illustrated in the bottom of Figure~\ref{top_BT1}). BT2 is a high-voltage low-current power delivery approach that includes a 48/POL converter within each UD on the top side of the wafer. In BT2, Cu pins are not used in the structure (less heat dissipation is expected due to the low-current nature of this topology), and the wafer is directly connected to a dedicated PCB using ball grid arrays (BGAs). Given that each UD in the BT2 topology, serves as a power source for surrounding FDs, the FD-UD tile configuration must be determined. Two tile configurations for BT2 are considered, namely, BT2\_8 and BT2\_24. BT2\_8 represents an 8-1 tile structure (shown in Figure~\ref{fig-coupon-8-1}), where each UD delivers power to 8 surrounding FDs. Alternatively, BT2\_24 represents a 24-1 tile structure (shown in Figure~\ref{fig-coupon-24-1}), where each UD delivers power to~24 nearby FDs. The size of the tile corresponds to a tradeoff between effective area and power losses. For example, assigning more FDs to each UD, \textit{i.e.}, larger tile, means more area is dedicated to FDs, leading to higher computing capacity on the one hand, and, on the other hand, to increased power loss and voltage noise. BT3 is a hybrid topology that includes a two-stage voltage conversion, a 48/12 converters on the dedicated external PCB and a 12/1 converters on the backside of the Si wafer.

A three-stage hierarchical decoupling capacitors (decaps) system is modeled for all power delivery topologies to allow for a fair comparison of voltage noise. In PT, decaps are placed on the topside of the wafer within the peripheral ring, in UDs, and in FDs. In BT1 and BT3, decaps are located on the external PCB, on the backside of the wafer, and on the top side of the wafer. For BT2, decaps are located on the external PCB, on the top side of the wafer, and also within the UDs. Details regarding the type and density of decaps are provided in Table~\ref{tab_decaps}. Furthermore, as typically practiced, a fourth stage of decaps can be included on the FDs for all topologies.

\begin{table*}[htbp]
	\vspace{-0.05in}
	\caption{Parameters of the considered decaps for the proposed power delivery topologies~\cite{CR13}.}
	\label{tab_decaps}
	\vspace{-0.05in}
	\renewcommand{\arraystretch}{1}
	{\begin{center}
			\begin{adjustbox}{width=2\columnwidth,center}
			\begin{tabular}{lccccc}
				\toprule
			\textbf{Location}    & \textbf{Type} & \textbf{Topologies}    & \textbf{Capacitor density (nF/mm\textsuperscript{2})} & \textbf{ESL (nH)} & \textbf{ESR (m$\Omega$) / time constant (ns)} 	\\																					  
				\toprule
                PCB & electrolytic capacitor& BT1, BT2, BT3  & 2,600 -- 4,800 & 3000 -- 6000 & 10 -- 20 \\
                Backside of the wafer & ceramic capacitor & BT1, BT3  & 8,800 -- 40,000& 300 -- 400 & 10 -- 20 \\
                On top of the wafer & DTCAP & PT, BT1, BT2, BT3  & 300 -- 1500 & Negligible & 2 < RC < 20\\
                Inside UDs & CMOS capacitors & PT, BT2  & 1 -- 3 & Negligible & RC < 250 \\
				In side FDs & CMOS capacitors & PT & 1 -- 3 & Negligible & RC < 250 \\
	
				\bottomrule
			\end{tabular}
			\end{adjustbox}
		\end{center}
	}
	\vspace{-0.2in}
\end{table*}
\section{Modeling Applications of Interest}
\label{section_3}
A wafer-scale platform supports the integration of a large number of cores as well as high memory capacity. Applications with highly parallel workloads are, therefore, the best candidates for utilizing such platforms \cite{CR10}. As such, AI accelerators, neuromorphic computing, and HPC have been selected as the applications of interest as well as their commercial representatives, respectively, Intel Loihi \cite{CR16}, AMD EPYC \cite{CR17}, and Google tensor processing unit (TPU) \cite{CR18}. Due to the lack of available data for newer generations, the specifications of the first generation of these three commercial products have been used. Given that dielets on the Si-IF are unpackaged chips, specifications of the selected products in the unpackaged form are required. These specifications are listed in Table~\ref{table_specs}. A dielet on the Si-IF should optimally be of an area in the range of 1--100~mm\textsuperscript{2}, based on tradeoffs among IP reusability, yield, testing complexity, handling considerations, and I/O complexity/power \cite{CR6}. Based on the parameters of functional blocks within the floorplan of each commercial application, including power delivery, memory bandwidth, processing capabilities, and communication protocols, the unpackaged chip of each application is divided into several dielets to meet the area requirement of the Si-IF platform and in line with the dielet paradigm shift in system integration \cite{CR29}. The specifications of the Si-IF-compatible dielets for each product are also listed in Table~\ref{table_specs}.

\begin{table*}[htbp]
	\caption{Specifications of the three commercial applications and the Si-IF-compatible dielet for each application.}
	\label{table_specs}
  	\vspace{-0.15in}
	\renewcommand{\arraystretch}{0.9}
	{\begin{center}
			\begin{tabular}{lllllll}

				\cmidrule[\heavyrulewidth](l){2-7}
				&\multicolumn{2}{c}{\textbf{Intel Loihi (neuromorphic computing)}}	&\multicolumn{2}{c}{\textbf{AMD EPYC (HPC)}} &\multicolumn{2}{c}{\textbf{Google TPU (AI accelerator)}}\\
				\cmidrule[\heavyrulewidth](lr){2-3} \cmidrule[\heavyrulewidth](l){4-5}
				\cmidrule[\heavyrulewidth](l){6-7}
				 &Original Chip & Si-IF dielet & Original Chip & Si-IF dielet &Original Chip & Si-IF dielet \\
				\toprule
				Platform & SoC & Si-IF & MCM & Si-IF & SoC & Si-IF\\
				Area (mm\textsuperscript{2})& 60 & 60 & 852 & 85 & 331 & 82\\
				Number of chip(s) & 131 & 1 & 4 & 1 & 1 & 1 \\
				Process technology (nm) & 14 & 14 & 14 & 14 & 28 & 28\\
				TDP (W) & 0.082 & 0.085 & 180 & 25 & 75 & 40\\
				Supply voltage(s) (V) & 0.5 -- 1.25& 1 & 0.8 -- 1.4 & 1& 1 -- 5 & 1.8 \\
				Current (A) & 0.065 -- 0.164 & 0.085 & 128 -- 225 & 25 & 15 -- 75 & 22.22 \\
				Frequency (GHz) & 0.032 & 0.032 & 2 -- 3 & 3 &0.7 & 0.7 \\
				Cores & 131\tmark[a] & 131 & 4$\times$ 8-core Chiplet& 4 & 1 & 1\\
				Throughput (TFLOPS) &1.26
				& 1.26 & 6.14 & 0.61 & 23 & 6 \\
				Memory bandwidth & 3.44 Gspike/s & 3.44 Gspike/s & 55~GB/s & 55~GB/s & 34~GB/s & 34~GB/s\\
				\bottomrule
				&&&&&&\\[-2ex]
				\multicolumn{7}{l}{$^a$~128$\times$(neuromorphic cores) + 3$\times$(x86 Quark cores)} \\
			\end{tabular}
		\end{center}
	}
	\vspace{-0.3in}
\end{table*}


SuperCHIPS is a short-range simple and low-power communication protocol for the Si-IF platform~\cite{CR19}. Based on the communication specifications among the functional blocks of each application and the communication parameters of SuperCHIPS~\cite{CR11}, a required power budget associated with short-range communication is added to the thermal design power (TDP) of each dielet. In other words, communication demand -- similar to the area, power,  and interconnect -- is adjusted for each commercial product to match the new requirements of the dielet assembly.
\section{Simulation Results and Discussion}
\label{sec_simulation_results}
Each commercial application was "reassembled" using Si-IF-compatible dielets and compatible electrical models were derived. The electrical model of the Si-IF structure is adopted from~\cite{CR9}. Decap values have been calculated based on expressions provided in~\cite{CR3}. For BT2, the electrical model of BGAs with 300~\textmu m diameter and 200~\textmu m height, is taken from~\cite{CR23}. For Intel Loihi, a low-power density system, the converter parameters are derived from~\cite{CR25}, whereas for the other two, high-power density, applications, from~\cite{CR24}. SPICE simulations were performed using MATLAB/Simulink~\cite{CR22}. 

\begin{figure*}[htbp]
    \vspace{0.2in}
    \centering
    \subfigure[]{
    \includegraphics[scale=0.33]{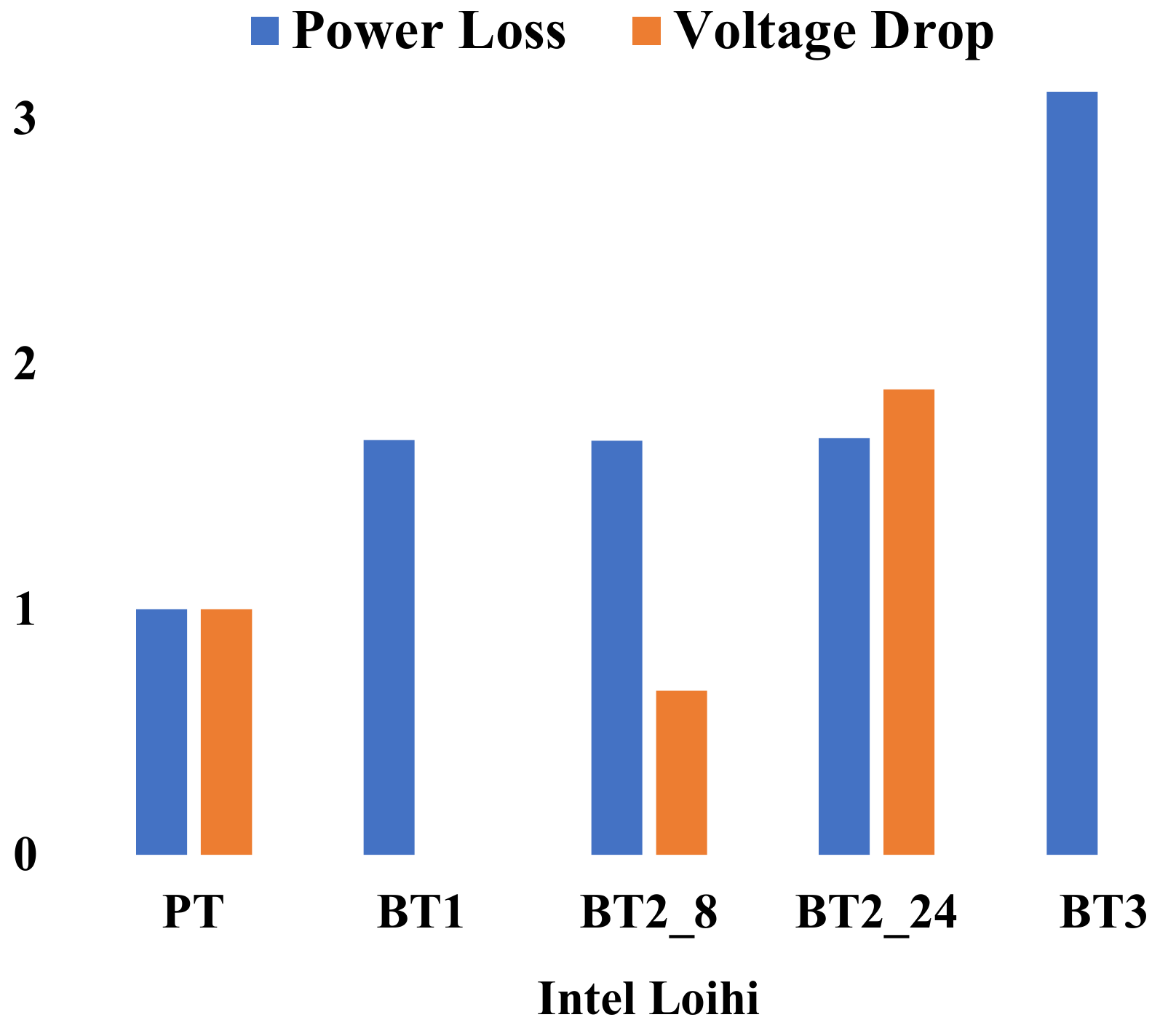}
    \label{chart_intel}
    }
    \subfigure[]{
    \includegraphics[scale=0.33]{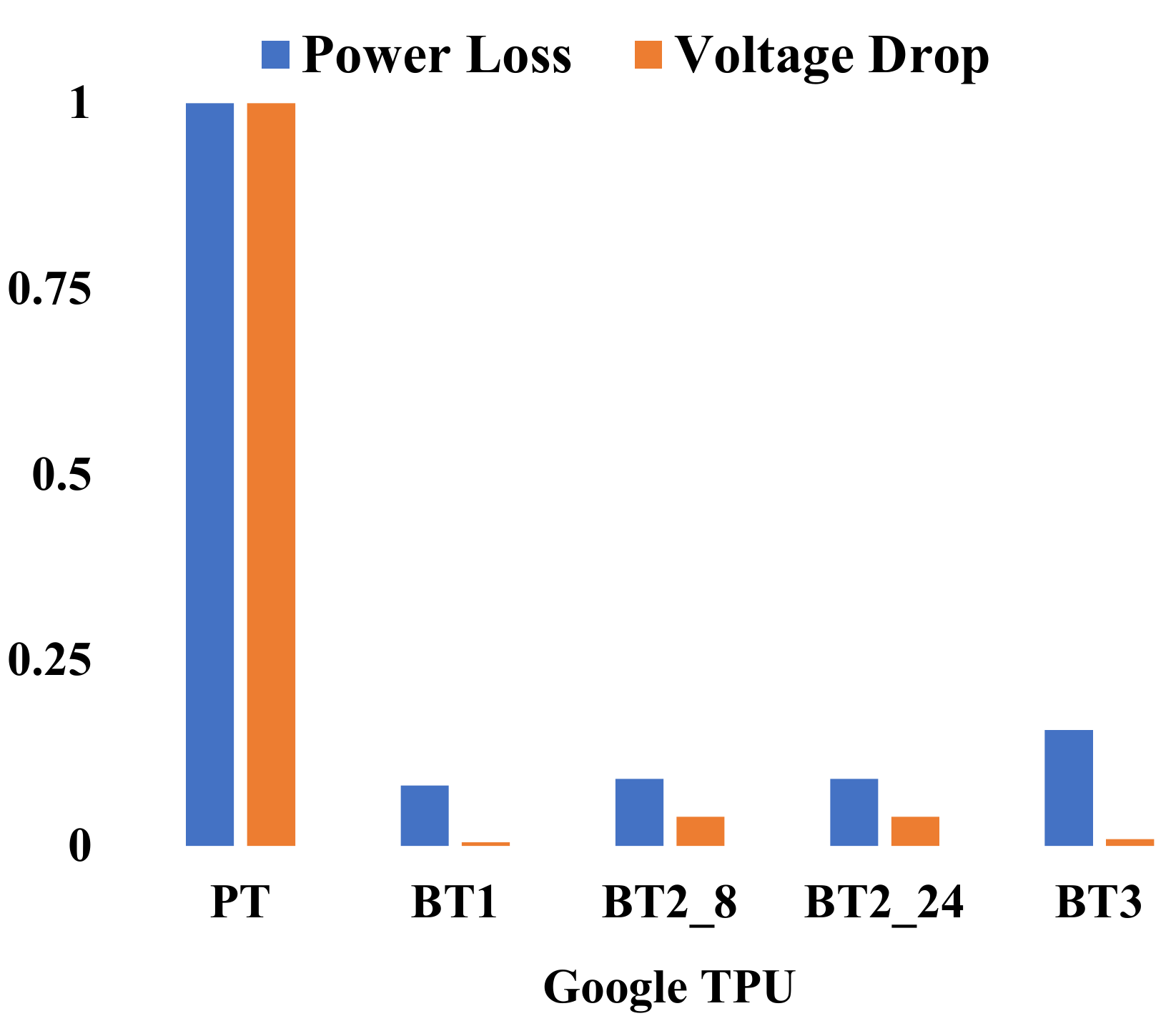}
    \label{chart_tpu}
    }
    \subfigure[]{
    \includegraphics[scale=0.33]{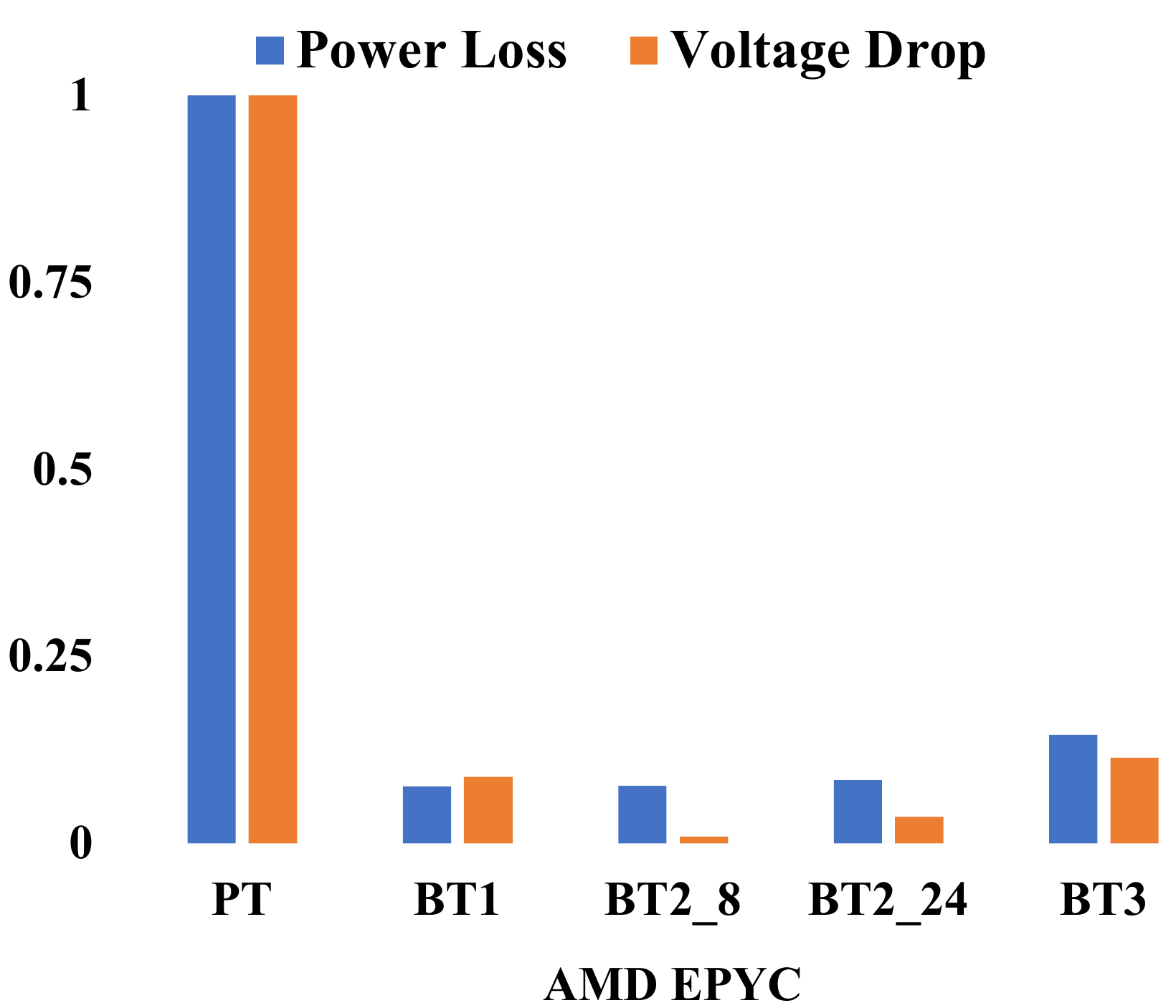}
    \label{chart_epyc}
    }

    \vspace{-0.1in}
    \caption{Simulated power loss and voltage drop of the three commercial applications on the Si-IF: (a) Intel Loihi (b) Google TPU (c) AMD EPYC. Note that the voltage drop values for BT1 and BT3 in (a) are about two orders of magnitude greater than the other values and therefore omitted from the plot for visibility.} 
    \label{sim_charts}     
\end{figure*}

The simulation results are shown in Figure~\ref{sim_charts}, demonstrating the total voltage drop (resistive voltage drop and inductive noise) and total power loss (resistive and inductive loss). All results are normalized to the values obtained for PT. 

It can be concluded from the results in Figure~\ref{chart_intel}, that for the scaled out Intel Loihi on the Si-IF, representing the neuromorphic computing application, PT is the best power delivery topology. From Figure~\ref{chart_tpu}, it can be concluded that for the scaled out Google TPU on the Si-IF, representing the AI accelerator application, backside power delivery has an advantage as compared to PT, and BT1, exhibiting a marginal superiority as compared to BT2, is the best power delivery topology. Finally, from the results shown in Figure~\ref{chart_epyc}, for the scaled out AMD EPYC on the Si-IF, representing the HPC application, BT2\_8, exhibiting a marginal superiority over BT2\_24, is the best power delivery topology. 

To provide a comprehensive comparison among the proposed power delivery topologies for the different applications, additional FOMs are considered. Specifically, a comparison of the applications in terms of area, computing performance, and power consumption are listed in Table~\ref{tab_perform}.
\begin{table*}[t]
	\caption{Comparison of FOMs of the three commercial applications "reassembled" on the Si-IF platform.}
	\label{tab_perform}
  	\vspace{-0.15in}
	\renewcommand{\arraystretch}{0.9}
	\setlength{\extrarowheight}{0.5pt}
	{\begin{center}
			\begin{adjustbox}{width=2\columnwidth,center}
			\begin{tabular}{clllllllllllllll}

				\cmidrule[\heavyrulewidth](l){2-16}
				&\multicolumn{5}{c}{\textbf{Intel Loihi on Si-IF}}	&\multicolumn{5}{c}{\textbf{AMD EPYC on Si-IF}} &\multicolumn{5}{c}{\textbf{Google TPU on Si-IF}}\\
				\cmidrule[\heavyrulewidth](lr){2-6} \cmidrule[\heavyrulewidth](l){7-11}
				\cmidrule[\heavyrulewidth](l){12-16}
				 &PT&BT1&BT2\_8& BT2\_24&BT3&PT&BT1&BT2\_8& BT2\_24&BT3&PT&BT1&BT2\_8& BT2\_24 &BT3  \\
				\toprule
            Number of FDs& 825& 1,178 &1,047 &1,131 &1,178 &582 &831 &739 &798 &831 &603 &862 &766 &828 &862 \\
            Total delivered &\multirow{2}{*}{70.1} &\multirow{2}{*}{100.1} & \multirow{2}{*}{89} & \multirow{2}{*}{96.1} & \multirow{2}{*}{100.1} & \multirow{2}{*}{14,550} & \multirow{2}{*}{20,775} & \multirow{2}{*}{18,475} & \multirow{2}{*}{19,950} & \multirow{2}{*}{20,775} & \multirow{2}{*}{24,120} & \multirow{2}{*}{34,480} & \multirow{2}{*}{30,460} & \multirow{2}{*}{33,120} & \multirow{2}{*}{34,480} \\
            \multicolumn{1}{c}{power (W)} & & & & & & & & & & & &&&& \\
            Computing performance & \multirow{2}{*}{1.04} & \multirow{2}{*}{1.48} & \multirow{2}{*}{1.32} & \multirow{2}{*}{1.43} & \multirow{2}{*}{1.48} & \multirow{2}{*}{0.73} & \multirow{2}{*}{1.05} & \multirow{2}{*}{0.93} & \multirow{2}{*}{1.01} & \multirow{2}{*}{1.05} & \multirow{2}{*}{3.62} & \multirow{2}{*}{5.17} & \multirow{2}{*}{4.6} & \multirow{2}{*}{4.97} & \multirow{2}{*}{5.17} \\ 
            \multicolumn{1}{c}{(PFLOPS)} & & & & & & & & & & & &&&& \\
				\bottomrule
		
			\end{tabular}
				\end{adjustbox}
		\end{center}
	}
	\vspace{-0.25in}
\end{table*}
The computing capacity, in floating point operations per second (FLOPS), of the "reassembled" applications on Si-IF, is also provided in Table~\ref{tab_perform} to allow for performance comparison. It can be seen from Table~\ref{tab_perform} that topologies BT1 and BT3 exhibit the highest performance for all applications, this is due to the largest area dedicated to FDs. Alternatively, it can be seen from the simulated results that, for extremely low-power density applications such as Intel Loihi, peripheral power delivery provides the best tradeoff. This is due to the low current demand by the application that significantly reduces the resistive loss on the top side of the Si-IF. Furthermore, in terms of fabrication considerations, PT topology is simplest. Whereas, fabricating Cu pins, placing decaps and converters at the backside of the wafer, and designing integrated converters within UDs add complexity to the backside power delivery topologies, significantly increasing the total cost.

The proposed power delivery topologies support a wide range of power, from tens of watts to tens of kilowatts. Furthermore, the proposed topologies support the integration of dielets with a wide variety of power densities, from 0.001 to 0.5~W/mm\textsuperscript{2}.

\section{Conclusions}
\label{sec_conclusions}
Three potential wafer-scale computing applications are modeled, simulated, and compared in terms of power and performance capacity. State-of-the-art commercial representative applications, including Intel Loihi, AMD EPYC, and Google TPU have been characterized for integration on the Si-IF. Four power delivery topologies, supporting a wide range of power densities and total power, have been proposed and compared based on several FOMs. Overall, demonstrated by simulation results, backside power delivery topologies exhibit a strong effect of inductive loss (due to the large interconnects in the backside of the Si-IF). This renders the backside power delivery topologies less effective for low-power applications, as compared to the peripheral topology. Alternatively, for high-power applications, backside power delivery topologies are best due to the highly resistive interconnect on the top side of the Si-IF. The obtained results confirm that the proposed power delivery topologies support integration of heterogeneous scalable systems on the Si-IF with extremely high computing performance. 

\vspace{-0.1in}
\section*{Acknowledgment}
This work was supported in part by the Natural Sciences and Engineering Research Council of Canada (NSERC) Grant RGPIN-2021-02778, and in part by the Fonds de Recherche du Qu\'ebec - Nature et Technologies (FRQNT) Grant 2022-NC-300132, and by McGill University.

\balance
\bibliographystyle{myIEEEtran} 
\small
\bibliography{references}

\end{document}